\documentclass[aps,prl,twocolumn,groupedaddress,superscriptaddress,showpacs]{revtex4-1}
\usepackage{graphicx}
\usepackage{amsmath}
\usepackage{hyperref}
\hypersetup{colorlinks=true, citecolor=blue, urlcolor=blue, linkcolor=blue}

\newcommand {\SPS}    {Sn$_2$P$_2$S$_6$ }
\begin{document}

\title{Double hysteresis loops in proper uniaxial ferroelectrics}

\author{I.~Zamaraite}
\affiliation{Faculty of Physics, Vilnius University, Sauletekio 9, 10222 Vilnius, Lithuania}
\author{R.~Yevych}
\email{ryevych@gmail.com}
\affiliation{Institute for Solid State Physics and Chemistry, Uzhgorod University, Pidgirna Str. 46, Uzhgorod, 88000, Ukraine}
\author{A.~Dziaugys}
\email{andrius.dziaugys@ff.vu.lt}
\affiliation{Faculty of Physics, Vilnius University, Sauletekio 9, 10222 Vilnius, Lithuania}
\author{A.~Molnar}
\affiliation{Institute for Solid State Physics and Chemistry, Uzhgorod University, Pidgirna Str. 46, Uzhgorod, 88000, Ukraine}
\author{J.~Banys}
\affiliation{Faculty of Physics, Vilnius University, Sauletekio 9, 10222 Vilnius, Lithuania}
\author{S.~Svirskas}
\affiliation{Faculty of Physics, Vilnius University, Sauletekio 9, 10222 Vilnius, Lithuania}
\author{Yu.~Vysochanskii}
\affiliation{Institute for Solid State Physics and Chemistry, Uzhgorod University, Pidgirna Str. 46, Uzhgorod, 88000, Ukraine}

\date{\today}

\begin{abstract}
For the first time in a bulk proper uniaxial ferroelectrics, double antiferroelectric-like hysteresis loops have been observed in the case of \SPS crystal. The quantum anharmonic oscillator model was proposed for description of such polarization switching process. This phenomenon is related to three-well local potential of spontaneous polarization fluctuations at peculiar negative ratio of coupling constants which correspond to inter-site interaction in given sublattice and interaction between two sublattices of \SPS modeled crystal structure. Obtained data can be used for development of triple-level cell type memory technology.
\end{abstract}

\maketitle

With the development of modern information technology, humanity is needed in ever larger volumes of digital information storage. Due to the fact, that the technological norms for reducing the size of memory cells gradually approach the limitations imposed by physical laws, memory manufacturers are looking for ways out of this situation. One of the methods for increasing the density of information storage is the use of memory cells with several logical states. In such systems, several bits of information are stored in one cell. A vivid example of this approach is a multi-level cell and triple-level cell (TLC) type flash memory~\cite{tor1999}. However, flash-technology has a number of drawbacks~\cite{meen1014}, which leads to the search for new types of memory cells. One of the candidates for the role of the non-volatile universal memory of the future is ferroelectric memory (Fe-RAM)~\cite{Park}. As it turned out, in these memory cells it is also possible to store several information bits~\cite{Baudry}. However, the practical implementation of such systems is only at the level of theoretical calculations.

In this Letter we propose the use of a room temperature proper uniaxial ferroelectric-semiconductor \SPS as an active material of a ferroelectric memory cell. The three-well local potential for the spontaneous polarization fluctuations~\cite{rusch2007} allows one cell to store three bits of information. This makes it possible to create on its base not only memory cells, but also non-Boolean information systems.

In \SPS crystals the second order phase transition from paraelectric phase (P2$_1$/n) into ferroelectric one (Pn) occurs at $T_0\approx338$~K. At room temperature spontaneous polarization is oriented in (010) monoclinic symmetry plane near [100] direction~\cite{vysoch2006}. Origin of spontaneous polarization is related to Sn$^{2+}$ cations electron lone pair stereoactivity together with valence fluctuations $\mathrm{P}^{4+}+\mathrm{P}^{4+}\longleftrightarrow\mathrm{P}^{3+}+\mathrm{P}^{5+}$ inside (P$_2$S$_6$)$^{4-}$ anions~\cite{rusch2007,glukh2012,yevych20161} which in whole can be considered in frame of second order Jahn-Teller effect~\cite{ber2013}. Thermodynamics of this mixed "displacive - order/disorder" continuous transition can be described in the Blume-Emery-Griffiths model~\cite{blum1971,ekiz}, which consider possibility of thermal fluctuations between three values of pseudospins ("-1", "0", "+1") in local three-well potential. This model predicts possibility of metastable nonpolar states inside of ferroelectric phase at $T<T_0$. By Monte-Carlo simulations with effective Hamiltonian, which was constructed on basis of frozen phonon approximation, a high probability of zero value for pseudospins inside ferroelectric phase of \SPS crystal was found~\cite{rusch2007}. Experimentally, by piezoelectric microscopy, inclusions of nonpolar "paraelectric regions" at $T<T_0$ have been directly observed~\cite{Kiselev}.

It is naturally to expect possibility of unusual switching of spontaneous polarization by external electric field in case of proper ferroelectric \SPS by "two-step" manner --- from one side well of the local potential into central well, and further into another side well. The metastable central-well state at switching process can be reflected as appearance of double hysteresis loops.

In this paper we present the results of hysteresis loops investigations for \SPS ferroelectrics. It was found that double loops can be observed in wide temperature interval of ferroelectric phase. Conditions for observation of "ordinary" ferroelectric loops are determined. Also, influence of the crystal lattice defects, that appear at partial substitution of tin by lead, or sulfur by selenium have been investigated. It was found that defects destroy double hysteresis loops in \SPS.

The quantum anharmonic oscillator (QAO) model is developed for description of double hysteresis loops in proper uniaxial ferroelectrics. Such model is based on description of temperature-pressure phase diagram for \SPS ferroelectrics~\cite{yevych20161} with account of by pressure flattening of side wells in the local three-well potential. "Ferroelectric" and "antiferroelectric-like" solutions of developed model demonstrate possibility of "ordinary" and "double" hysteresis loops realization in \SPS ferroelectrics.

For experimental investigations, crystals were grown using Bridgeman method. The silver paste was used for electrodes at polar cuts of plate shape samples with typical dimensions about $3\times5\times5$~mm$^3$.
The ferroelectric hysteresis (P-E) loops were measured by aixACCT TF Analyzer 2000 ferroelectric measurement system (aixACCT Co., Aachen, Germany) equipped with a high-voltage amplifier. The bipolar triangular voltage wave was applied on the sample for measurement of polarization. The frequency of the signal was varied in the range of $3\div30$~Hz. The polarization was determined from the experimental data of electrical current flowing through the sample.
Dielectric characterization was performed with a HP4284 precision LCR-meter at temperatures from 400~K to 100~K during the cooling cycle at a rate of about 1~K/min and frequencies ranging from 20~Hz to 1~MHz. At similar cooling conditions, pyroelectric coefficient was determined.

\begin{figure}
 \includegraphics*[width=3.4in]{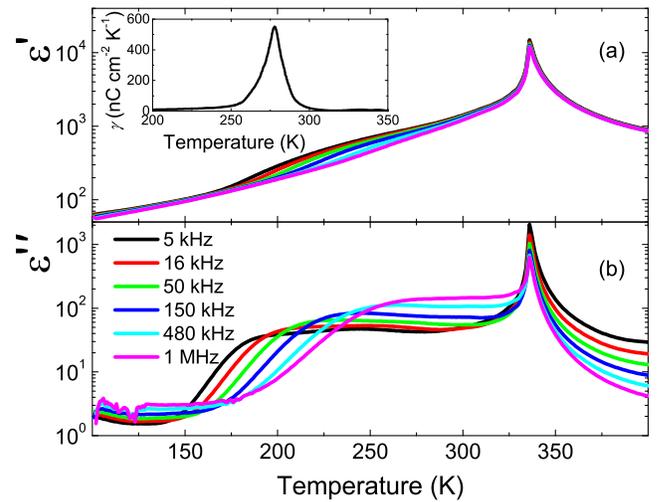}
 \caption{Temperature dependencies of (a) real and (b) imaginary parts of dielectric susceptibility for \SPS crystal at different frequencies of applied field. Temperature dependence of pyroelectric coefficient is shown in inset.\label{fig0}}
\end{figure}

As can be seen in Figure~\ref{fig0}, the second order phase transition is manifested by a sharp peak in dielectric susceptibility temperature dependencies near $T_0$ (for both real and imaginary parts). The dielectric losses in polar phase have enough high value but they rapidly decrease below 260~K. It is remarkable that temperature dependence of pyroelectric coefficient has't any anomaly near $T_0$, and demonstrate a wide temperature maxima near 270~K.

\begin{figure}
 \includegraphics*[width=3.4in]{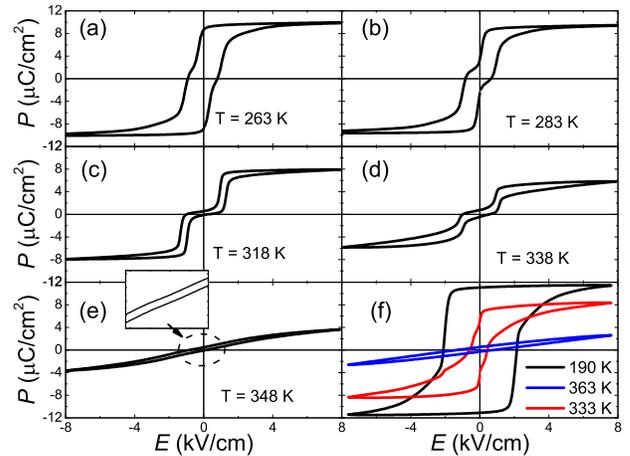}
 \caption{(a)-(e) The P-E hysteresis for as-grown \SPS sample at different temperatures. (f) The P-E hysteresis loops for \SPS sample, that was annealed in paraelectric phase at 383~K during 2 hours and after cooled till 190~K.\label{fig1}}
\end{figure}

For as-grown \SPS sample below $T_0$, the hysteresis loops have double-like shape (Fig.~\ref{fig1}(a)-(d)). At zero field, the spontaneous polarization is very small ($<1$~$\mu$C/cm$^{-2}$), but it reaches 8~$\mu$C/cm$^{-2}$ at increase of electric field amplitude to 6~kV/cm for $T=318$~K. At cooling below 283~K the hysteresis loops found usual ferroelectric shape with spontaneous polarization value near 10~$\mu$C/cm$^{-2}$ and coercitive field about 1~kV/cm. In paraelectric phase, above 338~K, the P-E loops found ellipse-like shape that is determined by electric conductivity of investigated sample.

After keeping of \SPS sample in paraelectric phase at temperature about 373~K during 2 hours and with further cooling to 190~K, ordinary ferroelectric loops have been observed (Fig.~\ref{fig1}(f)). At the same time, the spontaneous polarization and coercitive field reach its values about 11~$\mu\mathrm{C}\, \mathrm{cm}^{-2}$ and 2~kV/cm, respectively. But after keeping the sample at room temperature for a few weeks, antiferroelectric-like double loops have been observed again.
\begin{figure}
 \includegraphics*[width=3.in]{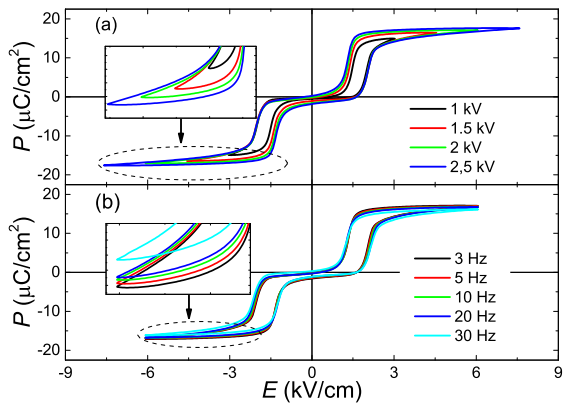}
 \caption{The P-E hysteresis loops measured in \SPS sample at room temperature with various switching parameters: (a) different voltage amplitudes on frequency 3~Hz; (b) different frequencies with 2~kV of voltage amplitude. \label{fig3}}
\end{figure}

The P-E hysteresis have been investigated at different switching parameters (frequency and electric field). In all measurements, the double hysteresis loops demonstrate their antiferroelectric-like shape in the whole ranges of change of parameters (Fig.~\ref{fig3}).

In (Pb$_y$Sn$_{1-y}$)$_2$P$_2$S$_6$ and Sn$_2$P$_2$(Se$_x$S$_{1-x}$)$_6$ mixed crystals, the usual ferroelectric loops have been observed. In the elementary cell of \SPS crystals, all atoms are placed in general positions~\cite{vysoch2006}. As follows, the atomic substitution increases crystal lattice defectives and avoids possibility of the double loops observation.

We suppose that observed peculiar switching in \SPS ferroelectrics is determined by local three-well potential for the spontaneous polarization fluctuations at peculiar negative ratio of inter-site interactions in given sublattice and between two sublattices of \SPS crystal structure~\cite{vysoch2006,rusch2007}. Below presented QAO model describes a possibility of double (antiferroelectric-like) hysteresis loops realization in such proper uniaxial ferroelectrics.

In this work we develop enhanced QAO model that was previously described for investigated crystal~\cite{yevych20161}. The main difference consists in model representation of a crystal lattice. The real crystal lattice was earlier described as one-dimensional chain of equivalent quantum anharmonic oscillators. The interaction of oscillators was described within the mean-field approach that makes possible to considerate such system as system of non-interacting oscillators. So, Hamiltonian of system had a form
\begin{equation}
  H=\sum_i\left(T(x_i)+V(x_i)+J\langle x\rangle x_i\right), \label{eq1}
\end{equation}
\noindent where $T(x_i)$, $V(x_i)$ are operators of kinetic and potential energy, respectively, $\langle x\rangle$ is an average value of displacement coordinate $x_i$, $J$ is an coupling constant. Here, the last term in Hamiltonian~(\ref{eq1}) reflects the mean-field approach by taking into account $\sum_{ij}J_{ij}x_ix_j\approx\sum_iJ\langle x\rangle x_i$ relation.

\begin{figure}
 \includegraphics*[width=3.in]{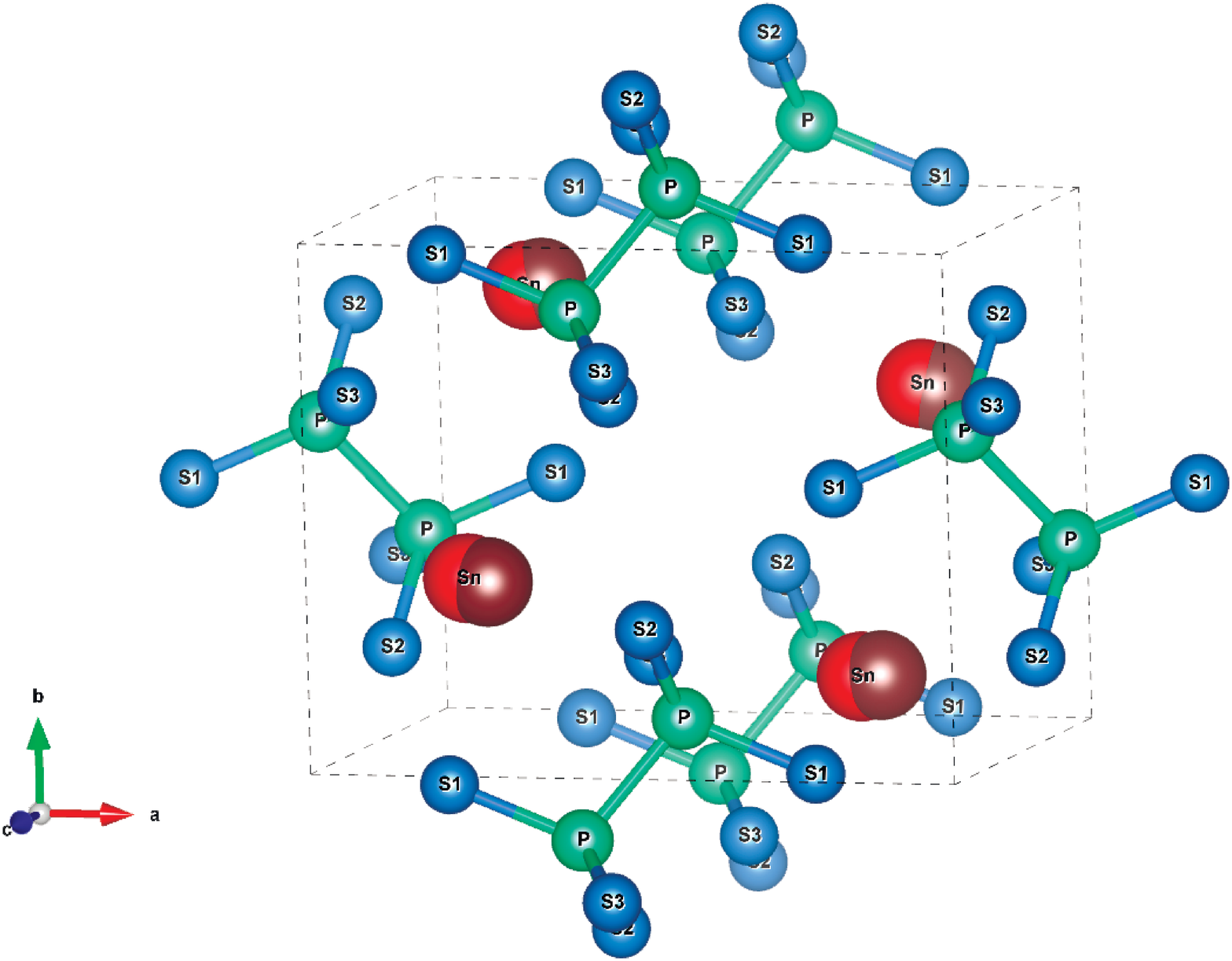}
 \includegraphics*[width=1.7in]{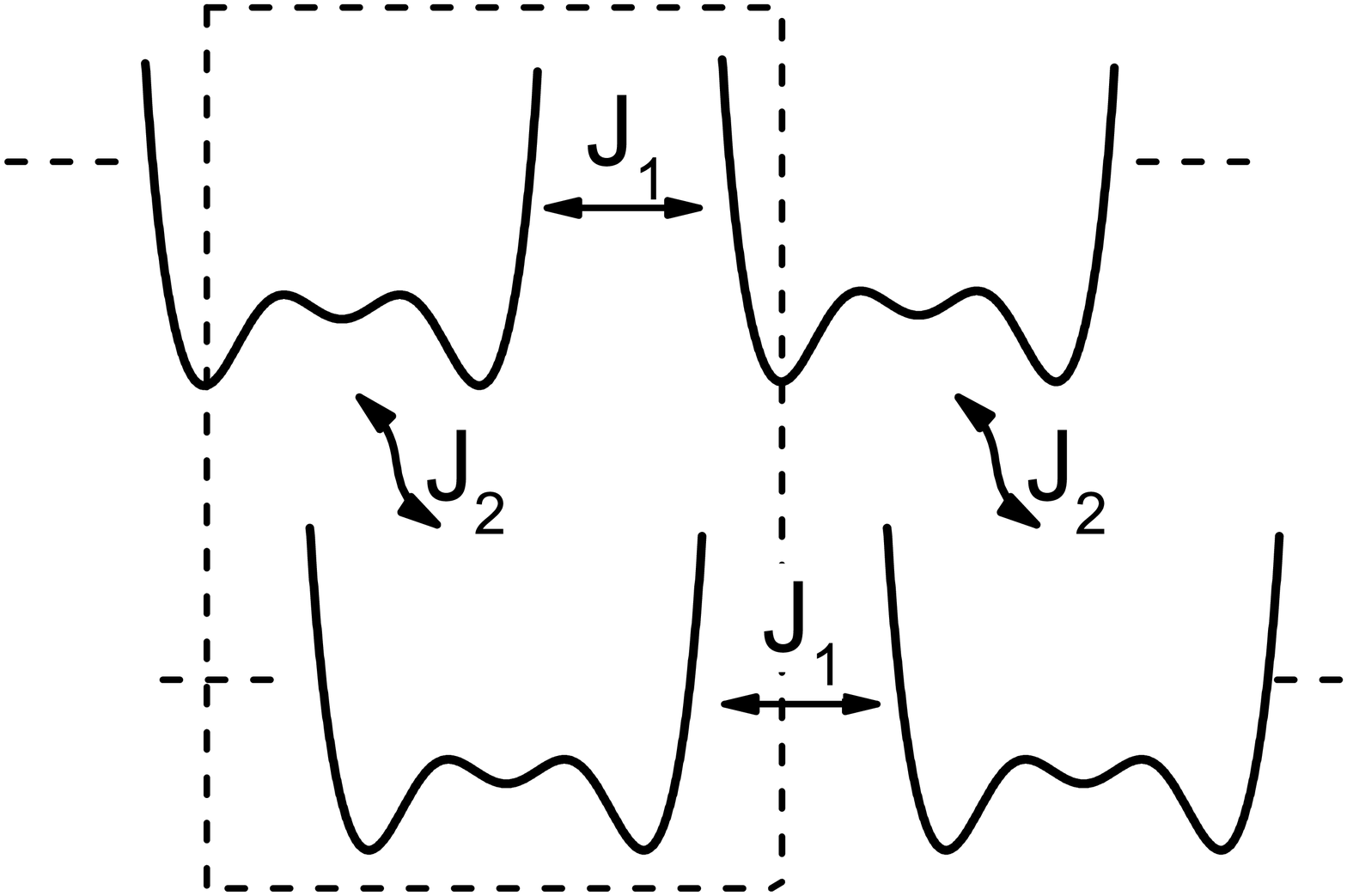}%
 \caption{(top) Crystal structure of \SPS crystal. Positions of tin cations (large balls) in paraelectric and ferroelectric phases are shown by different color. Two sublattices are presented by two formula units in monoclinic elementary cell; (bottom) QAO model with two chains that belong to two sublattices in \SPS crystal structure. Local dipoles in every \SPS formula units fluctuate in in three-well potential. Interaction within sublattices (chains) is determined by $J_1$ constant, between sublattices (chains) --- by $J_2$ constant. \label{fig4}}
\end{figure}

Suppose we now describe the \SPS crystal lattice as two interacted subsystem of oscillators. The elementary cell of \SPS contains two formula units, that are related by symmetry plane (see Fig.~\ref{fig4}). Despite the three-dimensional structure of \SPS crystal, its lattice can be presented as two chains of equivalent sites with local three-well potential. For simplicity, the chains are entirely identical and are characterized by the same local potential $V$ for each sites, and coupling constant $J_1$. However, interaction between oscillators from different subsystems is characterized by another one coupling constant $J_2$. Mean-field approach has been also applied. So, total Hamiltonian $H_t$ for such system can be written as a sum of two Hamiltonians $H_1$ and $H_2$ for each $x$ and $y$ subsystems
\begin{eqnarray}
    H_t &=& H_1 + H_2,\nonumber\\
  H_1&=&\sum_i\left(T(x_i)+V(x_i)+(J_1\langle x\rangle + J_2\langle y\rangle)x_i\right),\label{eq3}\\
  H_2&=&\sum_i\left(T(y_i)+V(y_i)+(J_2\langle x\rangle + J_1\langle y\rangle)y_i\right). \label{eq4}
\end{eqnarray}

Solving self-consistently system of equations (\ref{eq3})-(\ref{eq4}), one can obtain energy levels and corresponding wave functions of the system (see Ref.~\onlinecite{yevych20161} for details, and references therein). To investigate an influence of external electric field $E$ on model system, the terms proportional to $Ex_i$ and $Ey_i$ should be added to equation (\ref{eq3}) and (\ref{eq4}), respectively. Then, order parameter, which is proportional to a sum of average displacement of oscillators, and other physical properties can be calculated for different temperatures.

It should be noted that the shape of local three-well potential $V$ is phenomenologically described by nonlinear interaction $A_gB_u^2$ of several low energy optic modes of $B_u$ symmetry with fully symmetrical $A_g$ modes~\cite{rusch2007}. The shifts of Sn$^{2+}$ cations relatively to anions (Fig.~\ref{fig4}) are accompanied by rechargement and changing in chemical bonds covalency determining origin of the local electric dipoles in every \SPS formula units. In the given site, the pseudospin fluctuates in such local three-well potential what can be related to Hubbard-Holshtein model for systems with electronic correlations at involving of phononic excitations~\cite{fab1999}. Such description can be projected on BEG model with three values of pseudospins ("-1", "0", "+1") and with two order parameters --- dipolar with $B_u$ symmetry and quadrupolar with $A_g$ symmetry~\cite{rice}.

At evaluation of the QAO model parameters, the local three-well potential which was determined by first principles calculations~\cite{rusch2007} was taken into account. The "inter-chain" coupling constant $J_1$ was chosen to fit the second order phase transition temperature $T_0\approx338$~K at normal pressure. It was also accounted that for \SPS crystal the ratio of interactions $J_2/J_1\approx-0.23$~\cite{vysoch2006}, what follows from analysis of $T-x$ diagram of Sn$_2$P$_2$(Se$_x$S$_{1-x}$)$_6$ mixed crystals in axial next-nearest-neighbor Ising (ANNNI) model~\cite{selke} that explains observed Lifshitz point at $x_{LP}\approx0.28$ and continuous transitions from paraelectric phase into incommensurate phase at $x>x_{LP}$.
\begin{figure}
 \includegraphics*[width=3.in]{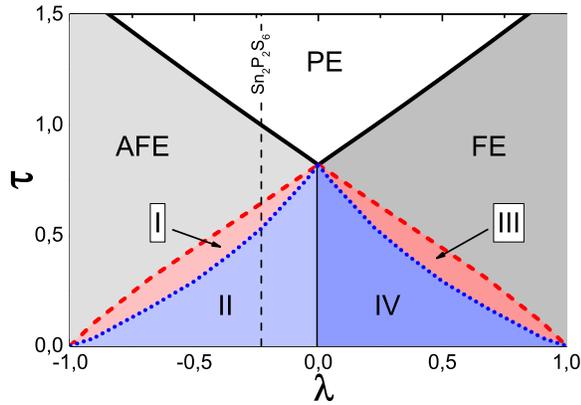}
 \caption{Reduced temperature-coupling constant ratio phase diagram for QAO model with stable paraelectric (PE), ferroelectric (FE) and antiferroelectric (AFE) phases. Coexistence of metastable AFE and FE states can be reached at temperature lowering: region I contains, in addition to stable AFE phase, one state with FE ordering, region II --- two FE states; region III contains, in addition to stable FE phase, one state with AFE ordering, region IV --- two AFE states. Thermodynamic path for \SPS crystal is pointed by vertical dashed line.\label{fig5}}
\end{figure}
\begin{figure}
 \includegraphics*[width=3.in]{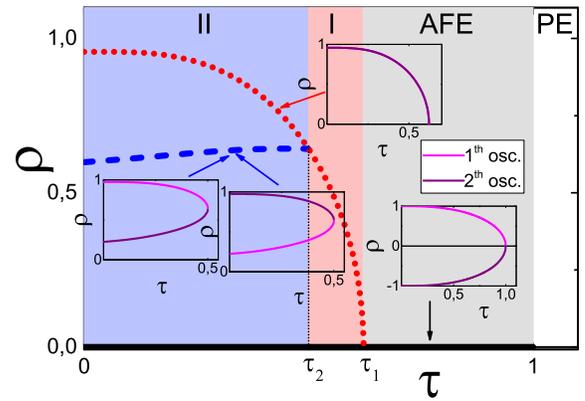}
 \caption{Temperature dependencies of total reduced order parameter for all solutions in the case of $\lambda=-0.23$. Thick black line corresponds to order parameter of AFE phase which continuously appear at $\tau\leq1$, red dotted line --- continuously appearance of FE phase at $\tau\leq\tau_1\approx0.62$, blue dashed line --- discontinuously appearance of FE phase at $\tau\leq\tau_2\approx0.5$. Magenta and purple lines on insets correspond to contributions into the total polarization from the first and the second subsystems, respectively. \label{fig6}}
\end{figure}

Let's introduce dimensionless parameters such as reduced temperature ($\tau=T/T_0$) and ratio of coupling constants $\lambda=J_2/J_1$. The calculated in proposed QAO model $\tau-\lambda$ phase diagram is shown in Figure~\ref{fig5}. It contains paraelectric (PE), ferroelectric (FE)  and antiferroelectric (AFE) phases, and four regions (I--IV) with metastable states. There are two points, $\lambda=1$ and $\lambda=-1$, where no metastable regions occur, and pure FE and AFE phases can be observed, respectively. Also, in the case of $\lambda=0$ (no interaction between different chains), FE or AFE phases can be realized with the equal probability. The $\lambda=-0.23$ point on phase diagram can be associated with model parameters for \SPS crystal. For such parameters, the AFE phase appears at cooling below $T_0$. But inside AFE phase, metastable FE states are also predicted (Fig.~\ref{fig5}). For this case, temperature dependencies of total reduced polarization $\rho$ (where $\rho=P/P_0$ and $P_0$ is a polarization in one subsystem at zero temperature for stable solution) for all solutions are presented on Figure~\ref{fig6}. Contributions from different chains to the total polarizations are shown on insets. As one can see, the phase transition from PE phase to AFE one at $\tau=1$ is continuous transition, and additional metastable states with FE ordering of oscillator displacements appear continuous at $\tau=\tau_1$ and discontinuous at $\tau=\tau_2$.
\begin{figure}
 \includegraphics*[width=3.4in]{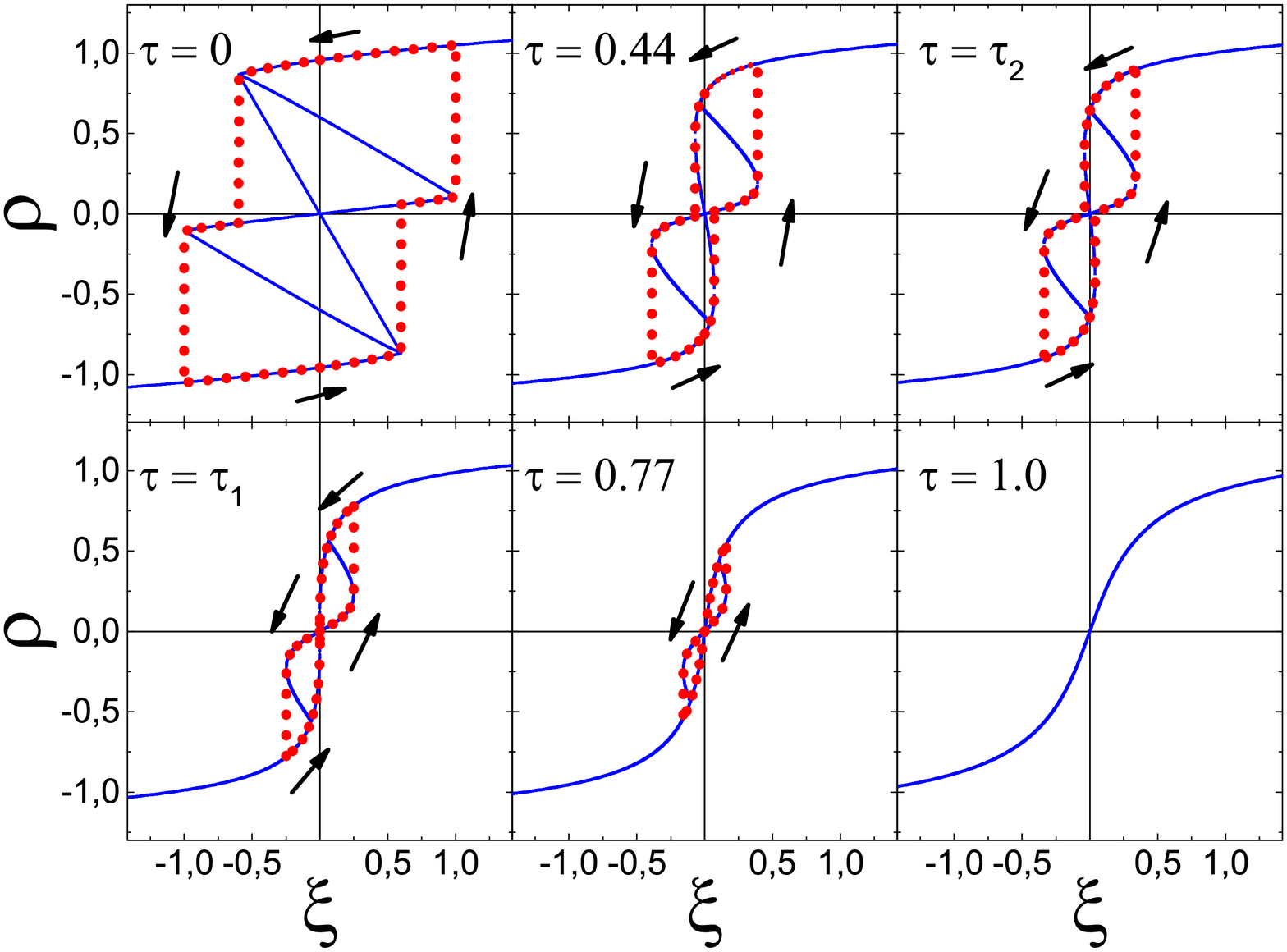}
 \caption{Calculated dependencies of polarization from external electric field at different temperatures for the case of $\lambda=-0.23$ (blue lines). Possible hysteresis loops for each temperatures are shown by red dotted lines.\label{fig7}}
\end{figure}

Calculated reduced polarization $\rho$ vs reduced electric field $\xi$ dependence (where $\xi=E/E_0$ and $E_0$ is a coercive field at zero temperature) for different temperature demonstrates several branches of stable and metastable solutions (Fig.~\ref{fig7}). At zero temperature on hysteresis loop, for example, there are three stable solutions, for which polarization increases with increasing of applied field, and three metastable solutions, for which polarization decreases with increasing of applied field. These solutions demonstrate a possibility of double hysteresis loop observation for proper uniaxial ferroelectrics \SPS with three-well local potential for spontaneous polarization fluctuations and at negative ratio of interactions inside and between structure sublattices.

Finally, the next interesting question --- why annealing in paraelectric phase destroy conditions for double loops observation at cooling below $T_0$? As we mentioned earlier, for \SPS ferroelectric-semiconductor the spontaneous polarization is related to tin cations stereoactivity and phosphorous cations valence fluctuations. Occurred recharging and changing in chemical bonds covalence can be presented as small hole polar formation at given $SnPS_3$ structural group and small electron polaron appearance in the nearest $SnPS_3$ structural group~\cite{yevych20161}.  Local dipole in \SPS formula units can be originated by polarinic exciton, and ferroelectric phase can be characterized as coherent state of polaronic excitons. At heating above $T_0$, free carriers are thermally excited. They can be trapped at the lattice imperfections producing dipole defects in paraelectric phase and, obviously, prevent observation of double dielectric loops below $T_0$.

In conclusion, polarization switching as double hysteresis loops in bulk proper uniaxial ferroelectrics was observed for the first time in the case of \SPS crystal. Such peculiarity is related to three-well shape of local potential for spontaneous polarization fluctuations what determines a possibility of metastable nonpolar regions existence below second order phase transition temperature $T_0\approx338$~K. The origin of spontaneous polarization is determined by changes of chemical bonding that can be presented as second order Jahn-Teller effect based on Sn$^{2+}$ cations electron lone pair stereoactivity and $\mathrm{P}^{4+}+\mathrm{P}^{4+}\longleftrightarrow\mathrm{P}^{3+}+\mathrm{P}^{5+}$ valence fluctuations of phosphorous cations. The enhanced quantum anharmonic oscillators model is proposed which consider negative ratio of interactions inside and between sublattices in \SPS crystal. This model can explain coexistence of ferroelectric and antiferroelectric hysteresis loops in \SPS crystals. This observation can be used for development of TLC type memory technology.

\end{document}